\documentclass[12pt,preprint]{aastex}

\def\AU{{\rm\, AU}}

\def\K{{\rm\,K}}
\def\yr{{\rm\,yr}}

\begin{document}

\shortauthors{Chiang, Fischer, and Thommes}
\shorttitle{Extrasolar Planetary Eccentricities}

\title{Excitation of Orbital Eccentricities of Extrasolar Planets by Repeated
Resonance Crossings}

\author{E.~I.~Chiang, D.~Fischer, \& E.~Thommes}

\affil{Center for Integrative Planetary Sciences\\
Astronomy Department\\
University of California at Berkeley\\
Berkeley, CA~94720, USA}

\email{echiang, fischer, ethommes@astron.berkeley.edu}

\begin{abstract}
Orbits of known extrasolar planets that are located
outside the tidal circularization regions of their parent
stars are often substantially eccentric.
By contrast, planetary orbits in our Solar
System are approximately circular, reflecting planet formation
within a nearly axisymmetric, circumsolar disk.
We propose that orbital eccentricities may
be generated by
divergent orbital migration of two planets
in a viscously accreting circumstellar disk.
The migration is divergent in the sense that the ratio of the
orbital period of the outer planet to that of the inner planet
grows. As the period ratio diverges,
the planets traverse, but are not captured into,
a series of mean-motion resonances
that amplify their orbital eccentricities
in rough inverse proportion to their masses.
Strong viscosity gradients
in protoplanetary disks offer a way to reconcile
the circular orbits of Solar System gas giants
with the eccentric orbits of currently known extrasolar planets.
\end{abstract}


\section{INTRODUCTION}
\label{intro}

\begin{figure}
\epsscale{0.6}
\plotone{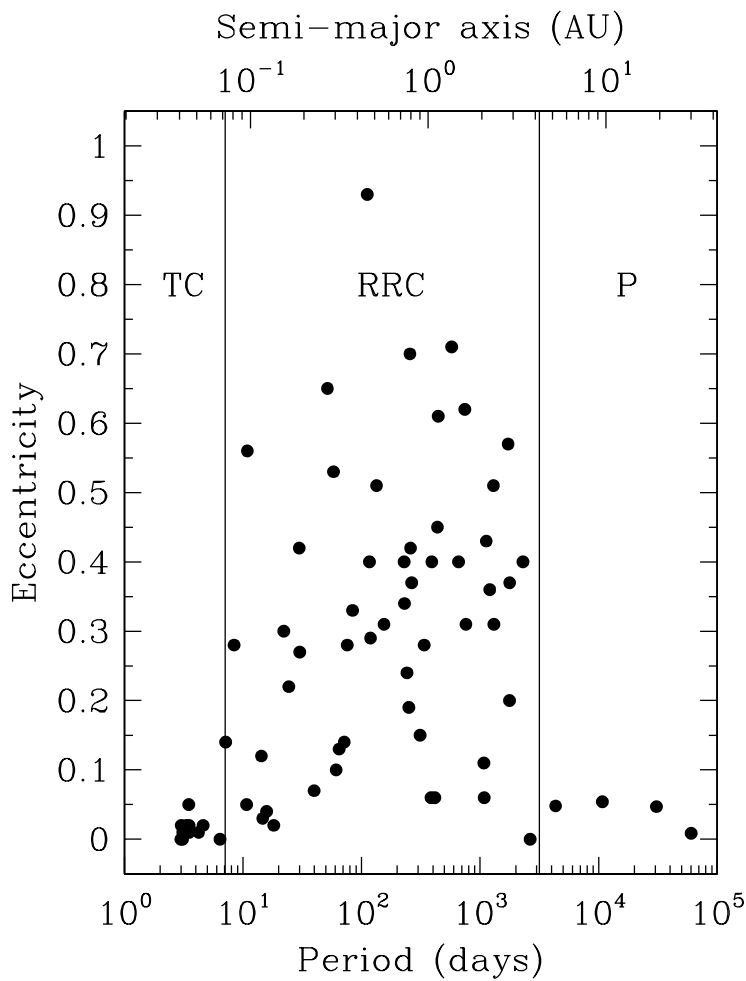}
\caption{Orbital eccentricities and periods of 64 planets.
The semi-major axis of the orbit is computed from the
period using a central stellar mass
of 1 $M_{\odot}$. TC represents the tidal circularization region, RRC the
regime proposed to have witnessed repeated resonance crossings,
and P the proposed outer passive region in which little or no migration
occurred.
\label{oper}}
\end{figure}

Orbital eccentricities $e$, periods $P$, and
semi-major axes $a$ of extrasolar planetary systems are plotted in Figure
\ref{oper}.
The rightmost four points at large $P$
represent the gas and ice giants in our Solar System (Lodders \& Fegler
1998). The leftmost cluster of points at small $P$
reflect tidal interactions between planets
and stars that erased whatever primordial eccentricities
these systems possessed (Lin et al.~2000). In the intermediate range
of periods, orbital eccentricities can be strikingly
large, typically exceeding those of giant planets
in the Solar System by factors of $\sim$2--20.

\placefigure{fig1}

Several theories have been proposed to explain
these large eccentricities.
Many encounter difficulties when applied
to the majority of systems, and one remains incompletely developed.
A stellar binary
companion can secularly drive a planet's eccentricity (Holman, Touma, \&
Tremaine 1997),
but nearly all known extrasolar
planets orbit solitary stars. Alternatively, dynamical
instabilities afflicting two planets
formed at close separation on
circular orbits can eject one planet while
inducing a large orbital eccentricity
in the remaining body (Rasio \& Ford 1996; Weidenschilling \& Marzari 1996).
However, close encounters
engineered in this fashion result also in planetary collisions,
leaving a large proportion of planets on circular orbits
that are not observed (Ford, Havlickova, \& Rasio 2001).
In a third scenario, orbital migration of two planets during which
the orbital period of one planet approaches that of the other (``convergent
migration'') can lead to resonant capture and eccentricity pumping.
While convergent migration and resonance capture likely underpin
the orbital eccentricities of
GJ 876b and GJ 876c, two planets observed to occupy a
2:1 resonance (Marcy et al.~2001b; Lee \& Peale 2001),
all other extrasolar planetary systems presently evince
no mean-motion resonant behavior.
A fourth explanation invokes gravitational interactions
between planets and the disks from which they formed (e.g., Artymowicz 1998).
The present theory of satellite-disk interactions has only been
derived to lowest order in $e$ (Goldreich \& Tremaine 1980);
it is inadequate when applied to extrasolar systems for which $e$'s
can be as large as 0.25--0.95.
In the present theory, whether the disk damps or excites $e$
depends sensitively on the distribution of disk material
near the planet (Goldreich \& Tremaine 1980; Papaloizou, Nelson, \& Masset
2001). While this distribution is currently uncertain,
Lee \& Peale (2001) find that the disk must strongly damp
eccentricities to reproduce the orbital parameters
of GJ 876 and to avoid the fine-tuning problem
of having the epoch of resonant capture coincide
with the dissipation of the disk.

This Letter proposes a fifth mechanism for exciting orbital
eccentricities: repeated resonance crossings of two planets
migrating on divergent trajectories.
In \S\ref{dddd} we argue that divergent orbital migration
of two planets is more likely than convergent migration.
In \S\ref{res} we demonstrate how divergent migration may lead
to substantial eccentricity excitation. In \S\ref{conc} we highlight
the requirements and qualitative predictions of our theory
and areas for future work.

\section{Disk-Driven Divergent Drift}
\label{dddd}

Migration of planets can be driven by tidal interactions with their
natal gaseous disks. Known extrasolar planets have sufficiently
large masses ($M \gtrsim M_J$, where $M_J$
is the mass of Jupiter) that they clear
annular gaps in disk material about their orbits (Ward 1997).
A planet that opens a gap is thereafter slaved
to the viscous evolution of its host disk, and
undergoes so-called ``Type II'' drift (Ward 1997).
The disk and its embedded planet
at stellocentric distance $r$ slide towards the star on the viscous diffusion
timescale,

\begin{equation}
\label{drif}
t_D = r/|\dot{r}| \sim r^2 / \nu \sim r^2 / \alpha c_s h \sim 2 \times 10^{4}
\left( \frac{r}{1\AU} \right)^{1/2} \left( \frac{1000\K}{T} \right) \left(
\frac{10^{-3}}{\alpha} \right) \yr \, .
\end{equation}

\noindent Here $\nu = \alpha c_s h$ is the
viscosity of the disk, $c_s$, $h$, and $T$ are the
sound speed, vertical scale height, and temperature of disk
gas, respectively, and $0 \leq \alpha \leq 1$
measures the strength of angular momentum transport
intrinsic to the disk. Equation (\ref{drif}) assumes
a central stellar mass $M_{\ast} = M_{\odot}$.

The diffusion time $t_D \propto \sqrt{r}/T\alpha$ almost certainly increases
with increasing $r$. Disk temperatures fall radially outwards
as the stellar flux and the disk's absolute gravitational
potential energy per unit mass diminish.
Sources of angular momentum transport include
(1) dissipation of density waves excited by numerous,
densely nested planets that are insufficiently
massive ($M \lesssim M_{\earth}$, where $M_{\earth}$
is the mass of the Earth) to open gaps (Goodman \& Rafikov 2001), and (2)
the magnetorotational instability (MRI) that
afflicts sufficiently ionized disks (e.g., Stone et al.~2000).
Mechanism 1 is capable of generating $\alpha \lesssim 10^{-3}$,
where the exact value depends on the spatial density
of small, as yet undetectable planets. Mechanism 2
has been demonstrated to generate $\alpha \sim 10^{-5}$--$10^{-1}$,
the exact value correlating positively with the electrical conductivity
of disk gas (Fleming, Stone, \& Hawley 2000).
The conductivity decreases with decreasing
temperature, so that under mechanism 2,
$d\alpha/dr < 0$.

The standard MRI
operates only in disk regions $r \lesssim r_d$ that are
sufficiently hot, $T \gtrsim 1000\K$,
that thermal ionization of trace metals
and sublimation of dust grains permit
the magnetic Reynolds number to exceed the
threshold required for instability.
Accretion disk models place $r_d$ between $\sim$0.1 AU
and $\sim$1 AU
(Gammie 1996; Bell et al.~1997; D'Alessio et al.~1998)---distances
at which many extrasolar planets are presently located (see Figure \ref{oper}).
In the absence of mechanism 1, it is possible
that $\alpha$ is effectively zero at $r \gtrsim 1 \AU$. While
Gammie (1996) has proposed that the standard MRI can still operate
wherever gas densities are sufficiently low that Galactic
cosmic rays can provide the requisite ionization levels,
the likelihood of this prospect remains unclear
for two reasons: (1) dust grains can severely reduce the electron
density, and (2) even neglecting dust, and even if the
magnetic Reynolds number exceeds the critical value
required for instability, the
time required for a neutral molecule to collide with an ion is typically
longer than a dynamical time, so that the bulk of the
mostly neutral gas fails to accrete with the ions
(Blaes \& Balbus 1994). We return to the possibility of a static outer disk
at the end of this Letter.

What is important for what follows are the two recognitions
that gap-opening planets at $r \lesssim$ a few AU
drift inwards at rates that are (1) extremely slow compared to
local orbital frequencies, and (2) different.
Since $dt_D/dr > 0$, two gap-opening planets at $r \lesssim$ a few AU
drift inwards such that the ratio of the period
of the outer planet, $P_2$, to that of the inner planet, $P_1$,
grows. The divergence of $P_2/P_1$ implies that a series
of mean-motion resonances will be crossed.
During each resonance crossing, the orbital
periods of the two bodies are momentarily commensurable;
that is, the ratio of their orbital periods approaches and then exceeds
a ratio of small, positive integers.

\section{Resonance Crossings on Divergent Orbits}
\label{res}

By contrast with the case where the period ratio
$P_2/P_1$ converges towards unity, the probability of resonant capture is zero
for diverging orbits (Sinclair 1972; Henrard \& Lemaitre 1983; Peale 1986; Yu
\& Tremaine 2001; and references therein).
Nonetheless, as in the convergent case, each divergent
passage can substantially alter the orbital
eccentricities (Henrard \& Lemaitre
1983; Peale 1986; Malhotra 1988; Dermott, Malhotra, \& Murray 1988)
and semi-major axes of the migrating bodies.

\placefigure{fig2}
\begin{figure}
\epsscale{0.6}
\plotone{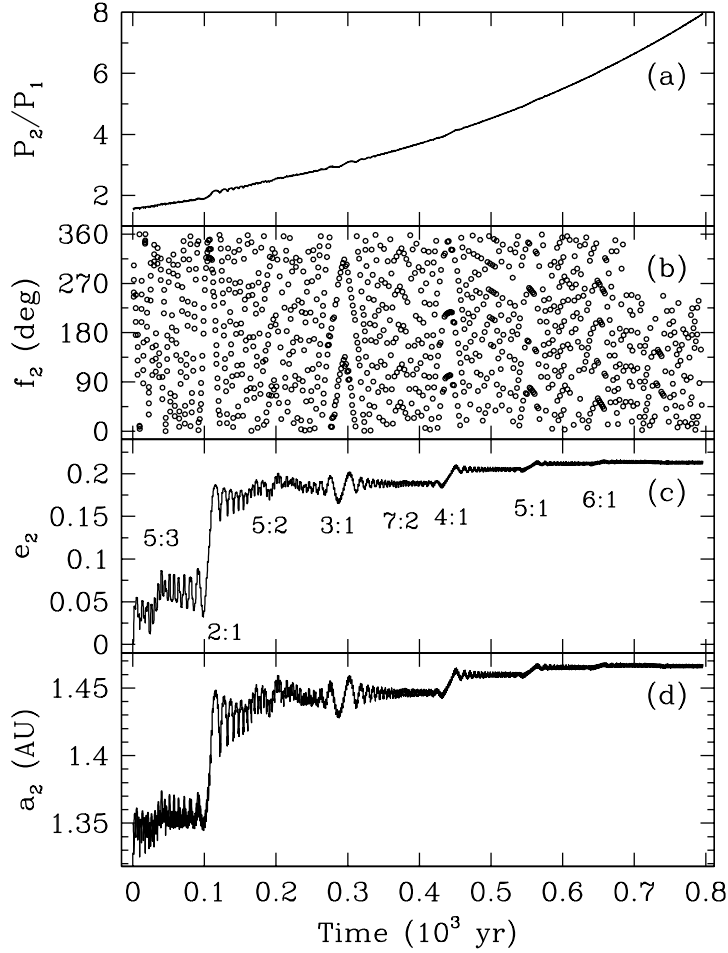}
\caption{Anatomy of resonance passages
involving an inwardly migrating massive planet
and a massless test particle on a more distant, co-planar orbit.
In panel (a) the ratio of orbital periods of the
test particle ($P_2$) and of the planet ($P_1$)
is plotted against time. Panel (b) plots the true anomaly
of the test particle whenever the particle comes
within 20$^{\circ}$ of the planet. Panels (c) and (d) plot the evolution of
the test particle's eccentricity and semi-major axis.
The resonances responsible for abrupt changes in the
eccentricity are labelled.
\label{anat}}
\end{figure}

We illustrate the underlying mechanics by considering the
problem of a massive, inwardly migrating planet
about a star, and its effect on a massless test particle on a more
distant, co-planar orbit. We take the mass of the planet $M_1$
to equal $1.5 \times 10^{-3} M_{\ast}$,
where $M_{\ast} = M_{\odot}$ is the mass of the
central star. Referenced to a coordinate system fixed on the star, the initial
orbital semi-major axis and eccentricity of the planet are
$a_1 = 1 \AU$ and $e_1 = 0$, respectively.
In addition to feeling the Newtonian force of gravity exerted by the star,
the planet feels an additional drag force
$\vec{F}_{drag} = -M_1 \vec{v}/t_{drag}$, where $\vec{v}$ is the
instantaneous velocity of the planet and $t_{drag} = 1.6 \times 10^3 \yr$
is the timescale over which $a_1$ decays to 0. This prescribed
drag force is introduced to simulate the effects of disk-induced
migration and does not directly affect the planet's eccentricity
(Papaloizou \& Larwood 2000).
Our value for $t_{drag}$ is chosen to illuminate
the evolution on timescales that are not too long compared
to $P_2$; larger values of $t_{drag}$
are probably more realistic and will be considered later.
The test particle is initially placed on an orbit having semi-major axis
$a_2 = 1.35$ AU and eccentricity $e_2 = 0$, and is initially
positioned at an angle $\Delta = 180^{\circ}$ away from the
angular position of the planet. The test particle
feels only the gravitational attraction from the planet and
from the star. The subsequent positions and velocities
of the planet and of the test particle are calculated using
a variable-order, variable-step Adams numerical integration scheme
(Hall \& Watt 1976).
Figure \ref{anat} displays the evolution. As the inner massive
planet migrates towards the star, the mean eccentricity of the
test particle undergoes 4 distinct changes. These changes
occur when the period ratio $P_2/P_1$ equals $5/3$, $2/1$,
$4/1$, and $5/1$. Concomitant changes in $a_2$
occur at these epochs of resonance passage. At times $t > 600 \yr$,
interactions between the particle and the now-distant planet
are negligibly small, and the particle is
left on a more eccentric ($e_2 = 0.21$) and slightly
expanded ($a_2 = 1.47$ AU) orbit compared to its initial one.

These changes in $e_2$ and $a_2$ occur because during passage
through a resonance, impulses of velocity are imparted to the test particle
from the planet at specific phases in the test particle's orbit
for extended periods of time.
Accelerations felt by the test particle from the inner massive
planet are strongest near times of conjunction, when positions of
the star, planet, and test particle fall on a straight line
and in that order. A $p:q$ resonance for which the planet executes integer $p$
circular orbits for every integer $q$ elliptical orbits traced
by the test particle is characterized by $|p-q|$ conjunctions
which occur at $|p-q|$ values of the particle's true
anomaly $f_2$.\footnote{True anomaly measures the angular
position of an object with respect to its periastron, and increases
in the direction of the object's motion.}
True anomalies $f_2$ near every conjunction
are plotted in Figure \ref{anat}b. For example, during passage through
the 2:1 resonance, conjunctions are repeatedly occurring
at values of $f_2$ concentrated in the interval
between $270^{\circ}$ and $360^{\circ}$. Conjunctions
in this quadrant amplify $e_2$ and $a_2$ (Murray \& Dermott 1999).
Analytic expressions for the magnitudes of eccentricity jumps through
first-order and second-order resonances are provided by Dermott, Malhotra, \&
Murray (1988).

\placefigure{fig3}
\begin{figure}
\epsscale{0.6}
\plotone{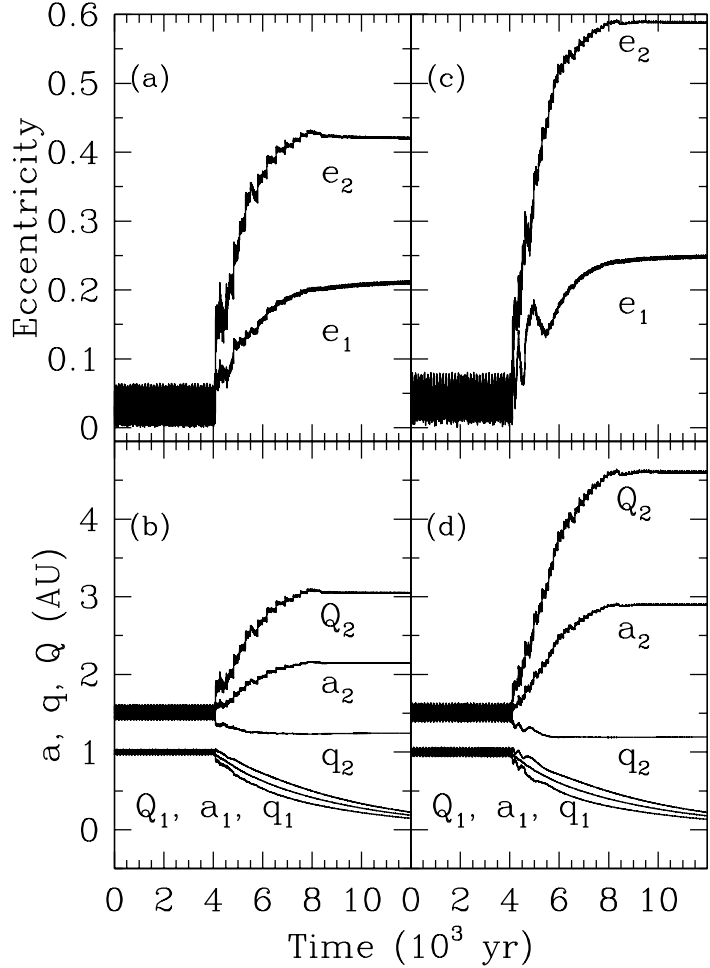}
\caption{Resonance passages involving two massive
planets. Masses of the inner and outer
planets are $M_1/M_{\ast} = 2\times 10^{-3}$
and $M_2/M_{\ast} = 1 \times 10^{-3}$, respectively.
The drag force is applied to the inner planet
over $t_{drag} = 1\times 10^4 \yr$
starting at $t_{start} = 4\times 10^{3}\yr$.
At $t = 0$, the ratio of semi-major axes
is $a_2/a_1 = 1.5$, the osculating eccentricities
are both 0.005 (ab) or 0.03 (cd), and the planets are
separated by an angle $\Delta = 180^{\circ}$.
At $t < t_{start}$, the
planets mutually excite eccentricities
of less than 0.075. Only after $t > t_{start}$
is differential migration introduced; the
eccentricities of both objects become
substantially excited through
repeated resonance crossings.
Panels (bd) also plot apastron
distances $Q = a (1+e)$ and periastron
distances $q = a (1-e)$.
\label{massi}}
\end{figure}

Accounting for the finite mass of the outer body
does not change our conclusions qualitatively.
In Figure \ref{massi} we showcase 2 scenarios involving
an inner body of mass $M_1/M_{\ast} = 2 \times 10^{-3}$
and an outer body of mass $M_2/M_{\ast} = 1 \times 10^{-3}$.
To demonstrate that the substantial eccentricities that are excited
in our simulation are caused by passages through resonances
and not by the mere close proximity of these massive bodies, we do not
impose any differential migration for the first $4\times 10^{3}\yr$
of the integration. The osculating eccentricities of both
bodies do not exceed 0.075 during this
phase when $\delta r$, the instantaneous distance between the
two planets, can be as small as 0.400 AU. Only when
the drag force is applied to the inner planet
at $t > 4\times 10^3\yr$, using $t_{drag} = 1 \times 10^4\yr$,
do the orbits diverge.
The eccentricities of outer and inner bodies then undergo resonant
excitation to values of $\sim$0.5 and $\sim$0.2, respectively,
in the step-wise fashion characteristic of resonance crossings.
The degree of amplification in $e_2$ exhibited in Figure \ref{massi}
is greater than that in Figure \ref{anat} primarily because
the former is based on a longer $t_{drag}$.
We have experimented with 20 different sets of initial orbital elements
and find that eccentricities $e > 0.2$ are excited
in all our test cases provided the initial $P_2/P_1 < 2$ so
that the 2:1 resonance is crossed.

The eccentricity and semi-major axis jumps are due
largely to resonant interaction and not to close encounters.
The separation $\delta r$ attains
minimum values of 0.325--0.400 AU (Figures \ref{massi}ab)
and 0.306--0.400 AU (Figures \ref{massi}cd) at $t = 4000$--$5000 \yr$,
whereas much of the amplification in $e_2$ and $a_2$
occurs over $t \approx 5000$--$6000 \yr$ when $\delta r > 0.400 \AU$.
We have verified by a separate integration
that freezing the divergent migration at $t = 4356 \yr$
in the case of Figures \ref{massi}cd---thereby freezing
$\delta r$ at its minimum value of 0.306 AU---does not generate
further eccentricity or semi-major axis jumps at subsequent times,
thus ruling out the significance of close encounters.

\section{Discussion}
\label{conc}

We have established that divergent orbital migration
of two planets can lead to significant
eccentricity excitation without resonance capture.
The effectiveness of this mechanism
hinges on the detailed, and here umodelled, interaction
between disk gas and the embedded planets. It is possible
that the disk gas exerts torques on the planets on
short enough timescales that the smooth navigation
of the resonant separatrix is disrupted; conjunctions
between the two planets may not occur at the same
orbital phases for extended periods of time due to
gravitational ``noise'' generated by disk gas.
This is a concern that we defer to future hydrodynamic
simulations of planet-disk interactions.

While we have staged our scenario within a gaseous circumstellar
disk, it seems possible that divergent migration
may also be effected by planetesimal scatterings
(e.g., Murray et al.~1998; Hahn \& Malhotra 1999).
In their simulations of the formation of the Oort Cloud
and the Kuiper Belt, Hahn \& Malhotra (1999) find that Uranus and Neptune
divergently migrate and cross a 2:1 resonance (see their Figure 6),
thereby undergoing eccentricity excitation.
The excitation is limited, however, because
the planetesimal masses which they
employ are large enough that the divergence
of the planetary orbits is not adiabatic.

Our proposed mechanism operates most effectively
when the initial orbits of the two bodies are sufficiently close that
powerful resonances for which $|p-q| = 1$ (e.g., the
2:1 resonance) are crossed during subsequent migration.
Formation of planets at such proximity
is not unreasonable; a single giant planet embedded within
a circumstellar disk may induce the collapse
of a second planet in the vicinity of the 2:1 resonance (Armitage \& Hansen
1999; Bryden et al.~2000).

Disk-driven divergent migration of two gap-opening planets is
most effective if a ring of viscous disk material remains present between the
two bodies. Kley (2000) and Bryden et al. (2000)
have performed pioneering numerical simulations
of two planets embedded in a disk resembling the minimum-mass
solar nebula. While these calculations have tentatively
shown that a ring can fail to be shepherded between
two planets, so that planetary orbits converge rather than diverge,
the outcome is model-dependent.
The results of Bryden et al. (2000) suggest that
if the mass of the ring is larger than the masses of the planets,
or if the ring's intrinsic $\alpha$ is
large so that planet-driven waves are efficiently dissipated near ring edges,
then the ring can be confined; see, in particular, their
model G.

The resonance crossings mechanism for generating eccentricities
demands that each planet have at least one other planetary companion,
either in the past or today. If $M_1 < M_2$, then $e_1 > e_2$, and vice
versa.
While the extrasolar planetary system $\upsilon$ And violates
this requirement (Butler et al.~1999; Chiang, Tabachnik, \& Tremaine 2001),
the system HD 168433 (Marcy et al.~2001a) satisfies it.
The absence of a second companion today in a given
planetary system may indicate
that the companion was accreted onto the star, either because
its eccentricity grew too large by resonance crossings,
or because ``Type II'' interactions drove the planet
to its fiery destruction. A signature of planetary
accretion would be an enhanced host star metallicity,
but enhancements due to accretion of Jupiter-like giants
during the first $\sim$$10^7 \yr$ of the life
of the star are too small to detect reliably (Murray et al.~2001).

The condition that the planets initially share the same orbit plane
may be relaxed. We would expect a non-zero initial mutual inclination,
$i$, to be amplified in analogous manner to the way eccentricities
are excited. Lifting the planet out of the plane
of the disk may represent a means of survival
against continued migration.
However, the degree of amplification
in $i$ is expected to be less than that in $e$ because
inclination resonances are at least second-order ($|p-q|\geq 2$)
in strength (Murray \& Dermott 1999).
Calculating the
relative orbital inclinations offers a means
of testing our theory; future astrometric missions
such as the Full-sky Astrometric Mapping Explorer (FAME)
and the Space Interferometry Mission (SIM), in conjunction
with stellar spectroscopic measurements, can place bounds on
the degree of misalignment between orbital axes of eccentric
planets and the spin axes of their parent stars.

If the MRI is the sole source of viscosity
late in the life of a protoplanetary disk,
we would expect gap-opening, Jupiter-mass planets
at distances outside a few AU to have suffered little to no
migration within the primordial gas disk. While it
is unconventional to think of T Tauri disks as
having $\alpha = 0$ at $r \gtrsim r_d \sim 1 \AU$, such a model
does not appear to violate observation or theory.
It would require the disk
to contain $\sim$0.01 $M_{\odot}$
inside $r \lesssim 1 \AU$, a condition for which the disk
remains gravitationally stable. Timescales
for accretion of this much material could be as long
as those observed, $\sim$$10^6\yr$, if $\alpha \sim 10^{-5}$
at $r \sim 1 \AU$.
Given a static outer disk, the theory of eccentricity excitation proposed here
would predict that orbits of giant planets at $r\gtrsim$ a
few AU be nearly circular. Giant planet orbits in our
Solar System conform to this expectation. We await the results
of ongoing Doppler velocity searches and
future space-based interferometric surveys
for extrasolar planets
to confirm whether the orbital architecture of the outer
Solar System is indeed commonplace.

\acknowledgements
We thank C.~McKee for thoughtful comments on the manuscript
and emboldening discussions, and P.~Goldreich,
R.~Malhotra, G.~Marcy, F.~Marzari, N.~Murray, E.~Quataert,
and S.~Tremaine for encouraging
and informative exchanges.

\end{document}